\documentclass[pra,twocolumn]{revtex4-1} 
\usepackage{graphicx}
\usepackage{amssymb, amsmath}
\usepackage{subfigure}

\newcommand{\micron}{\ensuremath{\mu\mathrm{m}}}

\begin{document} 
\date{\today}
\title{Discretely-observable continuous time quantum walks on M\"{o}bius strips and other exotic structures in 3D integrated photonics }
\author{M. Delanty and M. J. Steel}
\affiliation{MQ Photonics Research Centre and Centre for Ultrahigh bandwidth Devices for Optical Systems (CUDOS), Department of Physics and Astronomy, Macquarie University, Sydney, NSW 2109, Australia}

\begin{abstract}
We theoretically analyze the dynamical evolution of photonic quantum walks on M\"{o}bius strips and other exotic structures in 3D integrated photonics. Our flexible design allows discrete observations of continuous time quantum walks of photons in a variety of waveguide arrays.  Furthermore, our design allows one to inject photons during the evolution, allowing the possibility of interacting with the photons as they are `walking'.  We find that non-trivial array topologies introduce novel time-dependent symmetries of the two-photon correlations. These properties allow a large degree of control for quantum state engineering of multimode entangled states in these devices.
\end{abstract}

\maketitle

\section{Introduction}

With current advances in quantum technologies, quantum walks are beginning to be experimentally explored in a variety of systems.  The experimental study of quantum walks is of practical importance in  quantum simulation~\cite{QWQSimulation} and solving hard computational problems including the boson sampling problem~\cite{mqBosonSampling} and graph searches~\cite{GraphSearch}. Recently, quantum walks have been observed in waveguide arrays~\cite{OBrienQW,DTQW_Integrated,ThornPaper}, fiber loops~\cite{AutomatonPaper, MqSciencePaperRhode} and NMR~\cite{QWNMR}. Furthermore, quantum walks have been proposed in several other systems including ion traps~\cite{QWIon} and Rydberg atoms~\cite{QWRydberg}.

In particular, waveguide arrays in integrated photonics are a promising experimental implementation of quantum walks due to their flexible geometry and low losses.
Recent experiments in integrated photonics have demonstrated  a continuous time quantum walk (CTQW) in a 1D waveguide array ~\cite{OBrienQW} and a discrete time quantum walk in a quasi-1D array of directional couplers~\cite{DTQW_Integrated}.  A limitation of these approaches is that the length of the waveguide array is fixed and therefore it is only possible to observe the walk at a single time instant. Although it is possible, in principle, to reduce the array length by cut-back techniques, this `ad hoc' process is highly time consuming and achieving consistent coupling at each cut-back stage to allow meaningful comparison of photon correlations would be very tough.

Here, we describe an experimental proposal for a 3D waveguide array implementation of a CTQW, which can be observed at multiple discrete times (or discrete optical path lengths
depending on the reader's preferred mental picture). This generalizes previous waveguide approaches which could only observe CTQWs at a single time instant. The approach allows a variety of exotic structures with varying boundary conditions to be studied. Another novel aspect of our proposal is that photons can be added to the CTQW during its evolution. We show that by varying the waveguide boundary conditions and the time delay between the injection of photons, it is possible to engineer and analyze a range of multimode quantum states with time-dependent photon statistics. 

This paper is structured as follows, in section~\ref{SecDescription} we discuss our proposal for a general waveguide array. We then calculate the experimentally relevant two-photon correlation function for simultaneous and delayed two-photon input states in section~\ref{TwoPhotonInput}. We next analyze the dynamics of three particular examples of our proposal in section~\ref{SecExamples}, a cylindrical array, a M\"{o}bius strip waveguide array and a twisted circular waveguide array. In section~\ref{SecExperimentalIssues} we discuss practical aspects of an experimental implementation. We conclude in section~\ref{SecConclusion}.

\section{General Description of the Device \label{SecDescription}}

\begin{figure*}[t]
\centering
\subfigure[]{
{\includegraphics[scale=0.3]{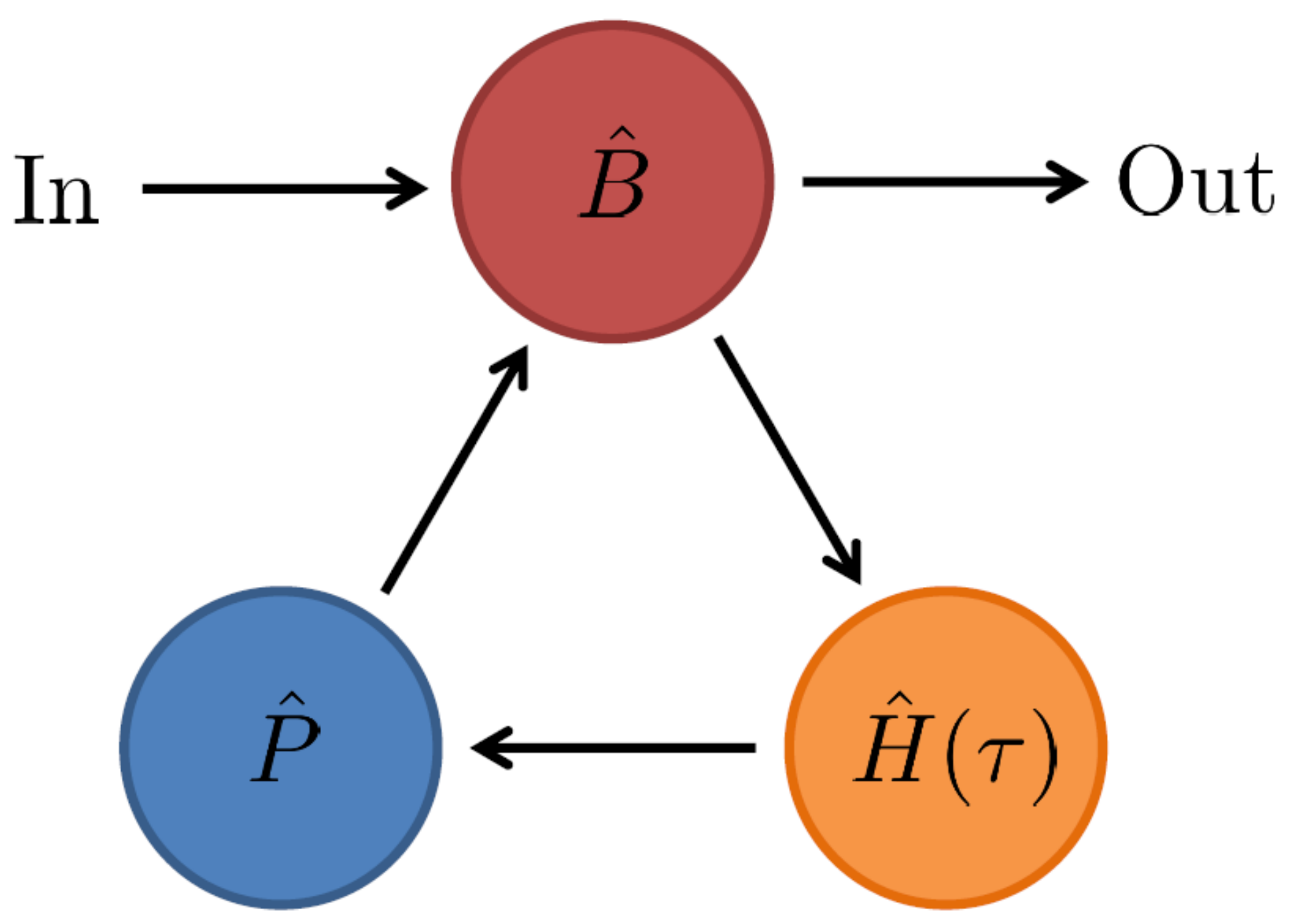}
\label{GeneralDiagram}
 }
}
\subfigure[]{
{\includegraphics[scale=0.25]{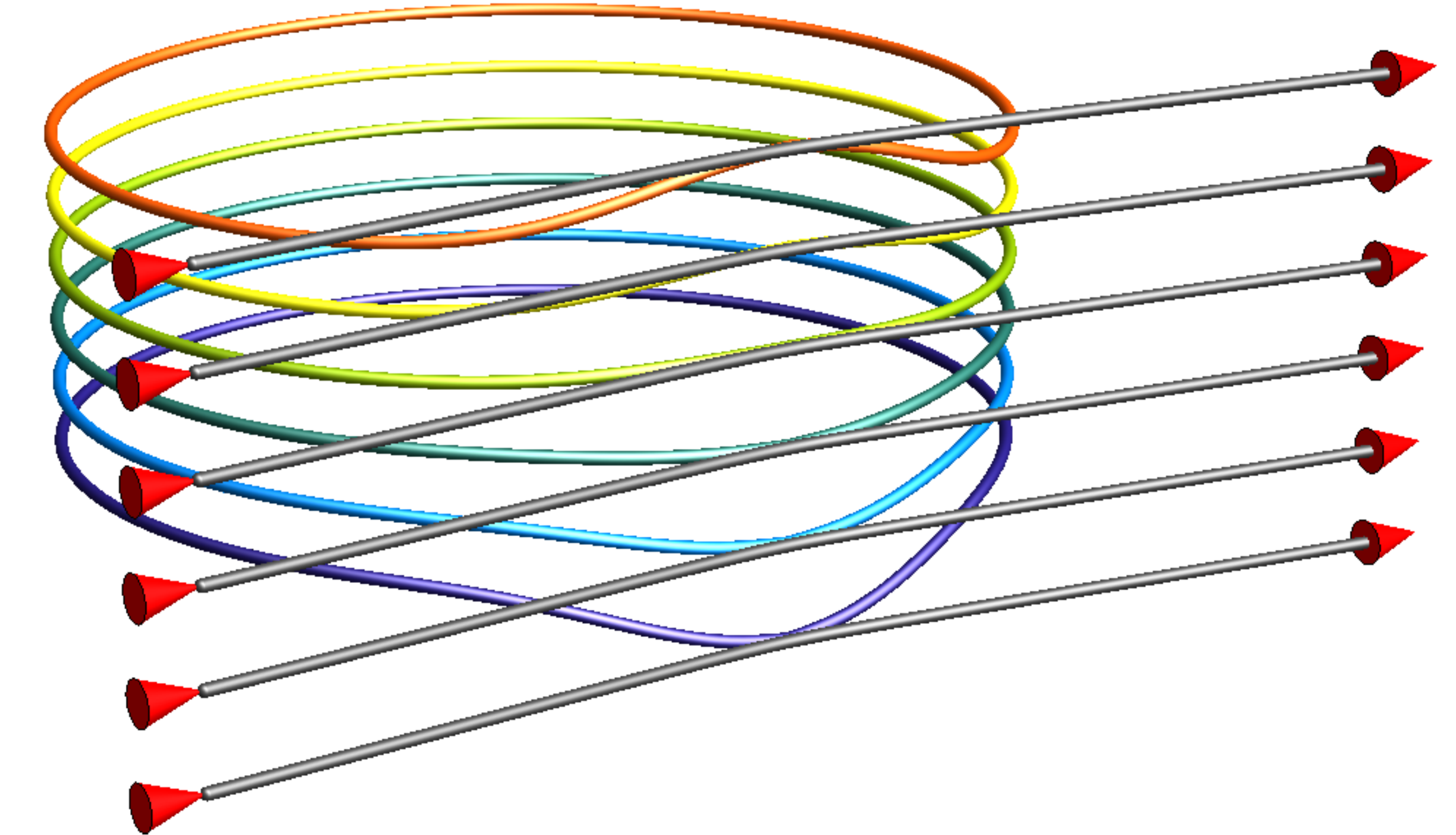}
\label{LinearCircleArray}
 }
}
\subfigure[]{
{\includegraphics[scale=0.3]{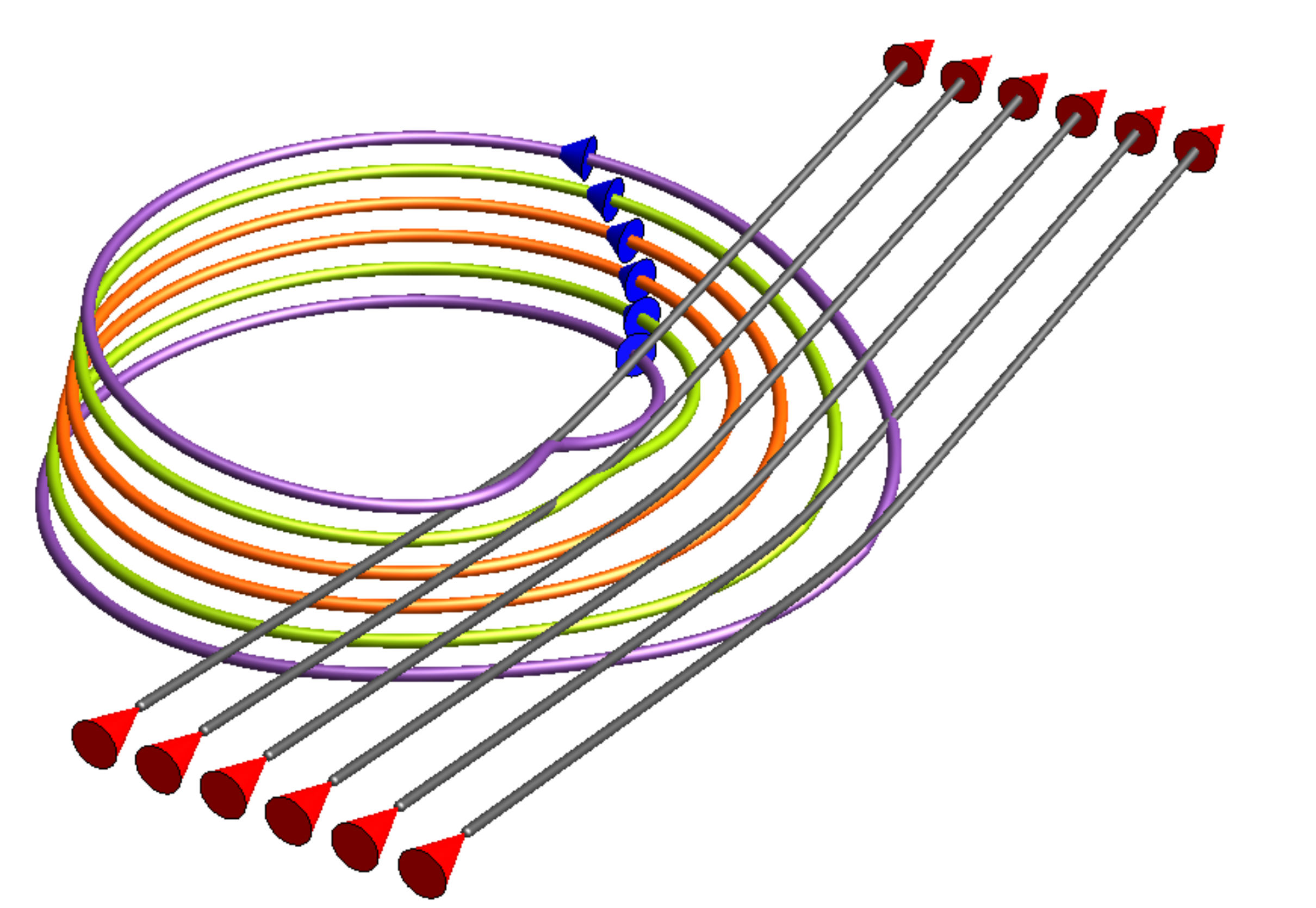} }
\label{Mobius_View1}
}
\subfigure[]{
{\includegraphics[scale=0.25]{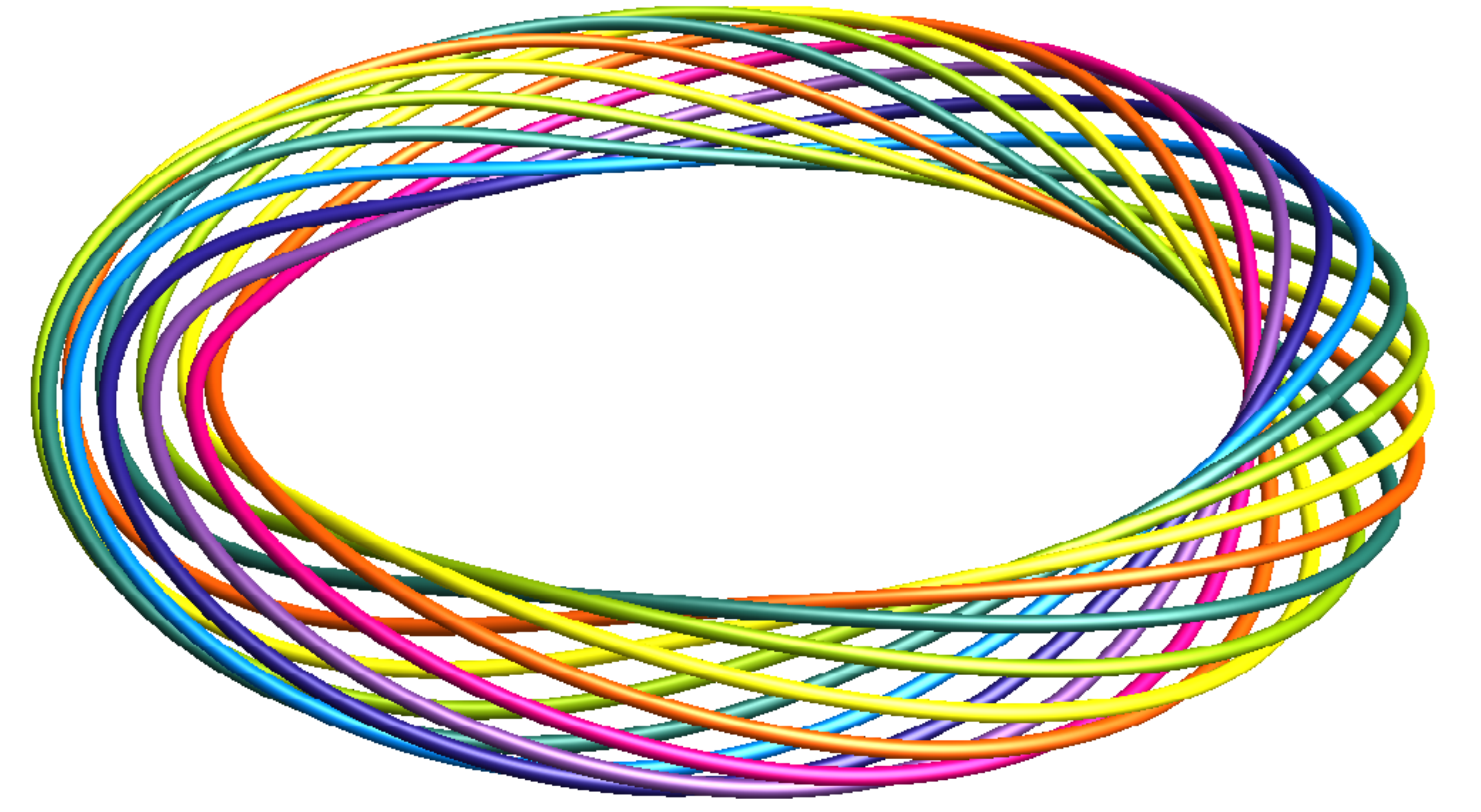} }
\label{Lines_of_Revolution_Circle}
}
\caption[]{ 
a) General schematic of experimental proposal. Photons are injected into directional couplers (denoted by $\hat{B}$), which couple into a waveguide array and evolve under a local Hamiltonian $\hat{H}$ for time $\tau$. Due to the geometry of the array, the waveguides then undergo a permutation denoted by $\hat{P}$. After the permutation the photons enter the directional couplers again and either exit the device to be detected or remain in the waveguide array for another loop.  b)--d) Three  specific implementation of the proposal: b) cylindrical array c) M\"{o}bius strip waveguide array and d) twisted circular waveguide array. For clarity the directional coupler array on the twisted circular array has been omitted.
}
\end{figure*}

We begin with  the dynamics of a general implementation of our waveguide array device.  We consider the conceptual device depicted in Fig.~\ref{GeneralDiagram}  with three examples in~\ref{LinearCircleArray}--(d). Initially, photons are injected into a set of directional couplers which either couple into a waveguide array or continue into a set of output modes to be detected. We then post-select events where all photons enter the waveguide array by monitoring the arrival times of the output photons.   We describe the meaning of the generic operators $\hat{B}$, $\hat{H}$ and $\hat{P}$ shortly.

Three particular examples of this general waveguide device are presented in Fig.~\ref{LinearCircleArray}--(d): the cylindrical array, the M\"{o}bius strip array and the twisted circular array. Each device has a radius much larger than the inter-waveguide distance and an array of directional couplers to couple the light in and out of the device. At the coupling region the waveguides fan-out so that each directional coupler only couples to a single waveguide. This region can also be used to correct any path length differences between the waveguides that is acquired during the loop.

The dynamics of the photons in the general waveguide array are determined by the  Hamiltonian,
\begin{eqnarray}\label{GeneralHam}
\hat{H} &=& \hbar \sum^{N}_{n,m=1} G_{n,m} \hat{a}^{\dagger}_n \hat{a}_m,
\end{eqnarray}
where, $G_{n,m}$ is the rate of coupling between the $n$-th and $m$-th waveguide. As the coupling matrix is real and Hermitian, $G_{n,m}=G_{m,n}$, it is possible to decompose it as $G = V \Lambda V^{\dagger}$, where $\Lambda_{i,j} \equiv \delta_{i,j} \lambda_j$ is a diagonal matrix of eigenvalues and the $j$-th column of the unitary matrix, $V$, is the $j$-th eigenvector of $G$. We denote the $n$-th element of the  $m$-th eigenvector of $G$ as  $ v_{n,m}$. Moving to the Heisenberg picture and using the eigendecomposition we can find the evolution of the operators~\cite{GlauberFock},
\begin{eqnarray}
\hat{a}_j (t) &= \sum^{N}_{k=1} U_{j,k} (t) \hat{a}_k (0), \label{HeisenbergA} 
\end{eqnarray}
 where
\begin{eqnarray}
U_{j,k} (t) &=  \sum^{N}_{p=1} e^{-i \lambda_p t} v_{j,p} v^*_{k,p}.\label{HeisenbergB}
\end{eqnarray}
These relations are sufficient to find the evolution of an arbitrary state in the waveguide array.

The input photons evolve in the waveguide array under the Hamiltonian (\ref{GeneralHam}) for time $\tau = L/v_g$, where $\tau$ is the time it takes for a single traversal of the waveguide array loop, $L$ is the path length of the loop and $v_g$ is the group velocity of photons in the array. (The structure is assumed to be
engineered such that the group velocity and path length of each loop are the same). After the system evolves for  $\tau$, due to the topology of the array the loop waveguide modes as labeled by the input/output couplers have encountered an effective permutation which we make explicit. 
For example, in the M\"{o}bius strip array depicted in Fig.~\ref{Mobius_View1} the permutation relabels the modes, $\hat{a}_j \rightarrow \hat{a}_{N+1-j}$, whereas for the twisted circular array in Fig.~\ref{Lines_of_Revolution_Circle} the permutation is $\hat{a}_j \rightarrow \hat{a}_{j+c}$ for some integer $c$. The cylindrical array in Fig.~\ref{LinearCircleArray} induces just the identity permutation. In general we denote the permutation by the operator $\hat{P}$ which has the action,
\begin{eqnarray}
\hat{a}_j \rightarrow \hat{a}_{\mathbf{p}(j)},
\label{Perms}
\end{eqnarray}
where $\mathbf{p}(j)$ is a bijective function of the mode number. We denote $n$ compositions of the permutation function as, $\mathbf{p}_n (j)$. For example, $\mathbf{p}_3 (j) = \mathbf{p}(\mathbf{p}(\mathbf{p}(j)))$, and similarly for  the inverse function, $\mathbf{p}^{-1}_3 (j) = \mathbf{p}^{-1}(\mathbf{p}^{-1}(\mathbf{p}^{-1}(j)))$ .

After the permutation, the photons encounter the directional couplers (the waveguide analog of a beamsplitter). The couplers act on the waveguides in the array ($\hat{a}_j$) and the input/output waveguides ($\hat{b}_j$) as
\begin{subequations}
\begin{eqnarray}
\hat{a}^{\dagger}_j &\rightarrow& \cos(\theta_j) \hat{a}^{\dagger}_j + i \sin(\theta_j) \hat{b}^{\dagger}_j,\\
\hat{b}^{\dagger}_j &\rightarrow&  i \sin(\theta_j)   \hat{a}^{\dagger}_j + \cos(\theta_j) \hat{b}^{\dagger}_j.
\end{eqnarray}
\label{bs}
\end{subequations}
With some probability (determined by $\theta_j$) a number of photons may exit the device to be detected. Alternatively, there is some probability that all photons will remain in the device. In this case, the photons will again evolve under $\hat{H}$ for time $\tau$, then undergo the permutation $\hat{P}$ before interacting once again with the directional couplers. Depending on the reflectivities of the couplers, all photons may remain in the waveguide array for a substantial period of time, $t\gg \tau$.

To summarize, photons enter the device via the input ports, then evolve under a local Hamiltonian; after this the waveguides undergo a permutation then interact with a set of directional couplers. Any photons that exit the device are measured and the photons remaining in the device repeat the process. The projected wavefunction after $n$ traversals of the device is
\begin{eqnarray}
|\psi (n \tau)\rangle  &=&\prod^{n}_{k=1}\hat{M} \hat{B} \hat{P} e^{-\frac{i}{\hbar} \hat{H} \tau }|\psi (0)\rangle ,
\label{psint}
\end{eqnarray}
where $\hat{B}$ denotes the action of the $N$ directional couplers and $\hat{M}$ denotes a projective measurement on the output $\hat{b}_j$ modes. 

The interaction with the directional couplers after each loop allows us to observe the Hamiltonian evolution at discrete time periods. Therefore the device can be used to discretely observe CTQWs of photons in a range of waveguide arrays. In addition, the permutation of the waveguides each step causes the device to exhibit interesting time-dependent symmetries of the photon statistics. By varying the initial state and selecting particular evolution times the device can be used to prepare a range of  multimode entangled output states with desired symmetry properties.

Finally we note that there is a class of states that are simultaneous eigenstates of both  $\hat{P}$ and $\hat{H}$. To find these states we first write the Hamiltonian as a sum of normal modes, $\hat{H} = \hbar \sum^{N}_{n=1} \lambda_n \hat{c}^{\dagger}_n \hat{c}_n$, where, $\hat{c}_j  = \sum^{N}_{k=1} v_{j,k} \hat{a}_k$. Eigenstates of $\hat{H}$ which are invariant under the permutation have the property $\hat{P}\hat{c}_j = \hat{c}_j$, from which we find the condition $v_{j, \mathbf{p}^{-1} (k) } = v_{j,k}$. In particular, two-photon simultaneous eigenstates have the interesting property that the observed two-photon correlation function is identical for all time steps, despite the permutation of the waveguides and the Hamiltonian evolution.

\section{Two-Photon Input \label{TwoPhotonInput}}

We now illustrate the evolution of photons in the device by giving two experimentally relevant examples. We first consider injecting two-photons into the device simultaneously for a CTQW and observing the time-dependent photon statistics. We also consider delaying the input of the second photon, which allows the experimentalist to interact with the walk as it evolves.

\subsection{Simultaneous Two-Photon Input}

Here we consider injecting a single photon into input waveguides $j$ and $k$ and post-selecting the events where both photons enter the waveguide array. The post-selection is achieved by rejecting events where a photon is detected in an output waveguide for $t< L/v_g$. If we further assume each directional coupler has the same angle $\theta$, the post-selected state occurs with probability $\sin^4 \theta$. The post-selected initial state of the waveguide array is therefore
\begin{eqnarray}
|\psi (0) \rangle &=& \hat{a}^{\dagger}_j \hat{a}^{\dagger}_k |\mathbf{0} \rangle,
\label{psijk}
\end{eqnarray}
where  the multimode vacuum is denoted $|\mathbf{0} \rangle = |0\rangle^{\otimes N}_a  |0\rangle^{\otimes N}_b$.

The state then evolves under the Hamiltonian $\hat{H}$ for time $\tau$. After this evolution and a permutation of the modes, the state becomes
\begin{eqnarray}
|\psi (\tau)\rangle &=& \hat{P} \hat{a}^{\dagger}_{j} (-\tau)
 \hat{a}^{\dagger}_{k} (-\tau) |\mathbf{0} \rangle\nonumber \\
 &=&\sum^{N}_{m,q=1}  U_{j,m} (\tau) U_{k,q} (\tau)  \hat{a}^{\dagger}_{\mathbf{p}(m)}
 \hat{a}^{\dagger}_{\mathbf{p}(q)} |\mathbf{0} \rangle, \nonumber
\end{eqnarray}
where the evolution of the modes, $\hat{a}^{\dagger}_{j} (-t)$, was found using (\ref{HeisenbergA}) and the relation (\ref{Perms}) was used to permute the modes.

This state then interacts with $N$ directional couplers according to (\ref{bs}),
\begin{eqnarray}
|\psi (\tau)\rangle  &=&\sum^{N}_{r,s=1}  U_{j,\mathbf{p}^{-1} (r)} (\tau) U_{k,\mathbf{p}^{-1} (s)} (\tau)
\nonumber\\
&\times& 
\left( \hat{a}^{\dagger}_r \cos \theta  + i \hat{b}^{\dagger}_r \sin \theta  \right)
\nonumber\\
&\times& \left( \hat{a}^{\dagger}_s \cos \theta  + i \hat{b}^{\dagger}_s \sin \theta   \right)|\mathbf{0} \rangle, \label{PsiTau}
\end{eqnarray}
where we have relabeled the summation indices to make explicit the permutations' effect on the coefficient of the mode operators. From this expression we find that at time $\tau$, the probability that both photons exit the device is $\sin^4 \theta$ and the probability that both photons remain in the device is $\cos^4 \theta$.

If both photons exit the device we can detect signatures of quantum interference using the two-photon correlation function, $\Gamma^{j,k}_{r,s} (t)$~\cite{OBrienQW, GlauberFock, SzameitYJunction}. This is defined as the probability that a single photon is detected in output waveguides $r$ and $s$ at time $t$, given that initially a single photon was injected into waveguides $j$ and $k$,
\begin{eqnarray}
\Gamma^{j,k}_{r,s} ( \tau) &=& \frac{1}{ 1 + \delta_{r,s}} \langle  \psi (\tau)|\hat{b}^{\dagger}_r (0) \hat{b}^{\dagger}_s (0) \hat{b}_s (0) \hat{b}_r (0) | \psi (\tau) \rangle.
\end{eqnarray}
  Using the wavefunction (\ref{PsiTau}), we find the two-photon correlation function after one traversal of the device is
\begin{widetext}
\begin{eqnarray}
\Gamma^{j,k}_{r,s} ( \tau) 
 &=&  \frac{\sin^4(\theta)}{ 1 + \delta_{r,s}} 
\left| U_{j,\mathbf{p}^{-1} (r)} (\tau) U_{k,\mathbf{p}^{-1} (s)} (\tau) + U_{j,\mathbf{p}^{-1} (s)} (\tau) U_{k,\mathbf{p}^{-1} (r)} (\tau) \right|^2. \label{Gamma1}
\end{eqnarray}
 We see that the two-photon correlation function is strongly dependent on the inverse permutation functions $\mathbf{p}^{-1} (r)$ and $\mathbf{p}^{-1} (s)$. Therefore the permutation is readily observable in the photon statistics. Furthermore, by comparing the two-photon correlation function to the classical correlation function for distinguishable particles~\cite{SzameitYJunction},
\begin{eqnarray}
P^{j,k}_{r,s} ( \tau) &=&  \frac{\sin^4(\theta)}{ 1 + \delta_{r,s}} \left(
\left| U_{j,\mathbf{p}^{-1} (r)} (\tau) U_{k,\mathbf{p}^{-1} (s)} (\tau)\right|^2 + \left| U_{j,\mathbf{p}^{-1} (s)} (\tau) U_{k,\mathbf{p}^{-1} (r)} (\tau) \right|^2 \right),
\end{eqnarray}
we see that quantum interference arises due to the fact that the two terms in (\ref{Gamma1}) are added prior to taking the modulus.

Alternatively, if both photons remain in the waveguide array, then we can determine their state at time $t=2 \tau$ by repeating the above steps of Hamiltonian evolution, permutation and the interaction with directional couplers. Continuing, for $n$ time steps, $t=n \tau$, we find the two-photon correlation function, 
\begin{eqnarray}
\Gamma^{j,k}_{r,s}  (n \tau) &=& \frac{\cos^{4(n-1)}(\theta)  \sin^4(\theta)}{ 1 + \delta_{r,s}}
\left| U_{j,\mathbf{p}^{-1}_n (r)} (n\tau) U_{k,\mathbf{p}^{-1}_n (s)} (n\tau) + U_{j,\mathbf{p}^{-1}_n (s)} (n\tau) U_{k,\mathbf{p}^{-1}_n (r)} (n\tau) \right|^2. \label{Gammat}
\end{eqnarray}
\end{widetext}
We see that the two-photon correlation function at $t=n \tau$ depends on $n$ applications of the inverse permutation function. This property leads to very interesting time-dependent photon statistics. For example for the M\"{o}bius strip considered in section~\ref{MobiusSection}, the permutation function has the property, $\mathbf{p}^{-1}_2 (s)=s$, and we find different statistics at odd and even time periods. By gating the output modes to only allow output at a particular time step, it is possible to prepare a range of multimode entangled states.

Another interesting property of the two-photon correlation function (\ref{Gammat}) is the pre-factor, $\cos^{4(n-1)}(\theta) \sin^4(\theta)$. This pre-factor reduces the  probability of a coincidence detection at later times, due to the interaction with the directional couplers at previous time steps. As choosing a weaker interaction with the directional couplers increases the probability that both photons remain in the device for longer \footnote{ We also note that it is more difficult to couple photons into a device that has a weak coupling to the directional couplers.  }, it is possible to optimize the coupling angle, $\theta$, to maximize the two-photon correlation function at time $t=n \tau$. For a desired time $t=n \tau$, we find that $\cos^{4(n-1)}(\theta) \sin^4(\theta)$ attains a maximum at $\cos(\theta) = \sqrt{n-1}/\sqrt{n}$. Therefore, the two-photon correlation function $\Gamma^{j,k}_{r,s}  (n \tau)$ is maximized  by choosing the optimal coupling angle, $\theta_\text{opt}= \cos^{-1}(\sqrt{n-1}/\sqrt{n})$. By building a variety of devices with differing coupling angles, it is possible to observe CTQWs over a range of  time periods.

\subsection{Two-Photon Input with Delay}

Our proposal also allows the possibility of injecting photons at different times. It is therefore possible to effectively interact with the quantum walk as it evolves. This novel feature is not possible in existing quantum walk experiments in integrated photonics.

Here, we consider injecting a photon into waveguide $j$ at $t=0$ and another into waveguide $k$ at $t=t_{d} \equiv n_d \tau$,  where $n_d$ is an integer.  We then monitor the output waveguides and post select events where both photons remain in the array for $t>t_d$. The wavefunction of the photons in the array immediately after the injection of the second photon is,
\begin{eqnarray}
|\psi ( t>t_d)\rangle &=& \hat{a}^{\dagger}_k (-(t-t_d)) \hat{a}^{\dagger}_{\mathbf{p}_{n_d} (j)} (-t)   |\mathbf{0} \rangle,
\end{eqnarray}
where the first photon has undergone $n_d$ permutations. We see that the second photon interacts with the permuted single photon quantum walk after it has evolved for $t_d$.

Repeating the analysis of the previous section we find the two-photon correlation function at  $t=n \tau +n_d \tau$ is
\begin{widetext}
\begin{eqnarray}
\Gamma^{j,k}_{r,s}  (n \tau +n_d \tau) &=&
\mathcal{N} \left| U_{j,\mathbf{p}^{-1}_{(n+n_d)} (r)} ((n+n_d)\tau) U_{k,\mathbf{p}^{-1}_n (s)} (n\tau) + 
U_{j,\mathbf{p}^{-1}_{(n+n_d)} (s)} ((n+n_d)\tau) U_{k,\mathbf{p}^{-1}_n (r)} (n\tau) \right|^2, \label{Gammatdelayed}
\end{eqnarray}
\end{widetext}
where $\mathcal{N}= \cos^{4(n-1)}(\theta)  \sin^4(\theta)/( 1 + \delta_{r,s})$. The  two-photon correlation function exhibits clear interference between the two walkers at different times. This property can create a range of entangled states, which can be found in a single experiment by varying the input ports and delay time $t_d$.

\section{Examples: Two-Photon input \label{SecExamples}}

\begin{figure*}[t]
\centering
\subfigure[]{
{\includegraphics[scale=0.7]{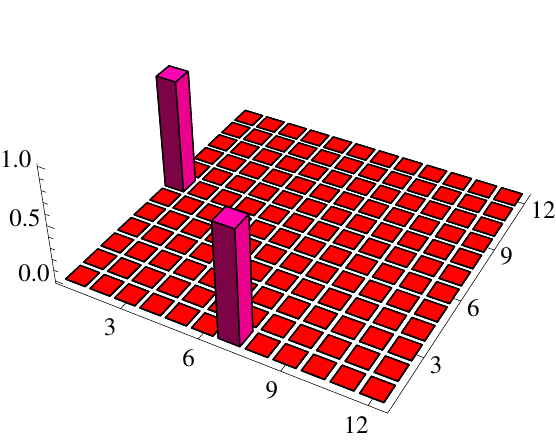}
\label{Corr_Linear_t0}
 }
}
\subfigure[]{
{\includegraphics[scale=0.7]{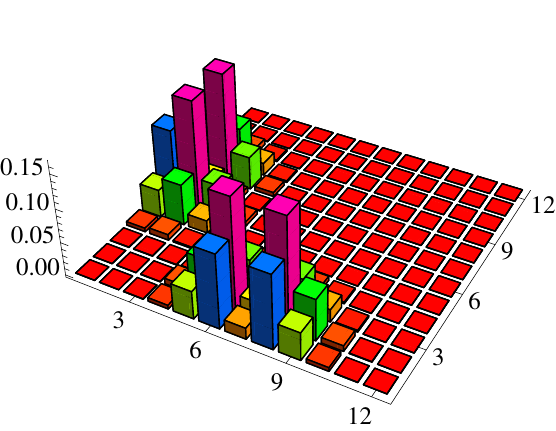} }
\label{Corr_Linear_t1}
}
\subfigure[]{
{\includegraphics[scale=0.7]{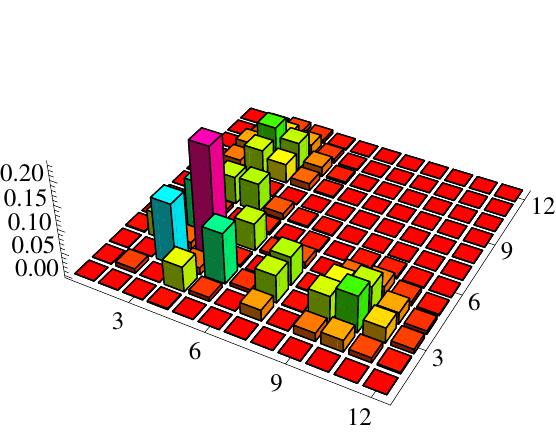} }
\label{Corr_Linear_t2}
}
\subfigure[]{
{\includegraphics[scale=0.7]{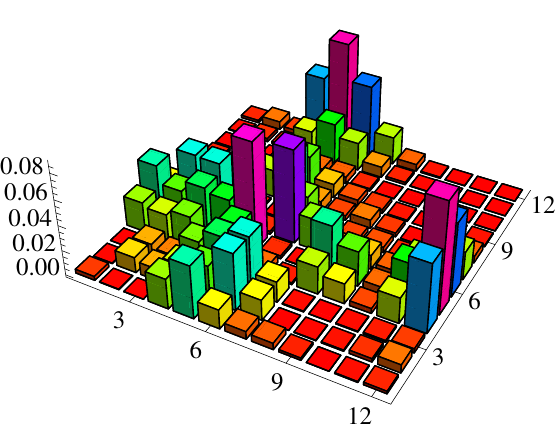} }
\label{Corr_Linear_t3}
}
\subfigure[]{
{\includegraphics[scale=0.7]{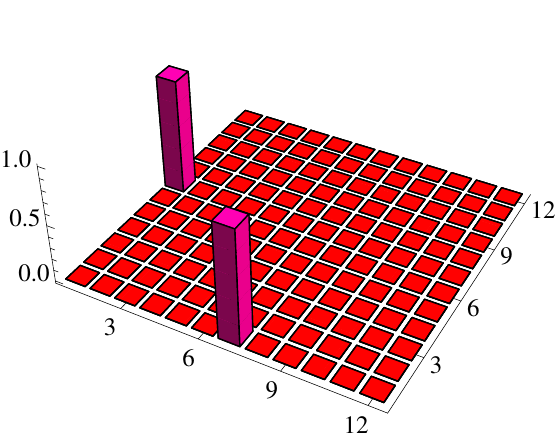}
\label{Corr_Mob_t0}
 }
}
\subfigure[]{
{\includegraphics[scale=0.7]{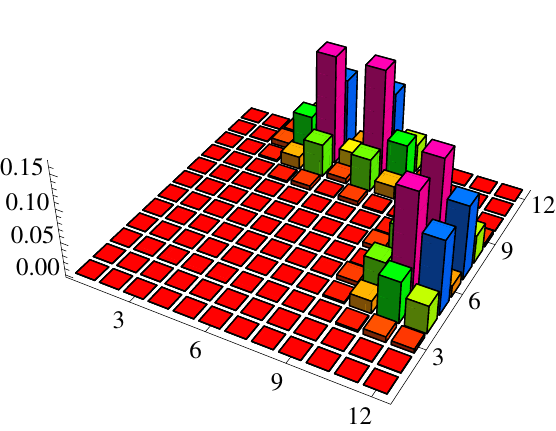} }
\label{Corr_Mob_t1}
}
\subfigure[]{
{\includegraphics[scale=0.7]{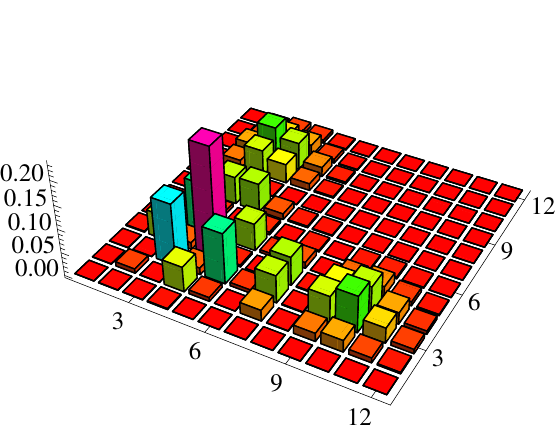} }
\label{Corr_Mob_t2}
}
\subfigure[]{
{\includegraphics[scale=0.7]{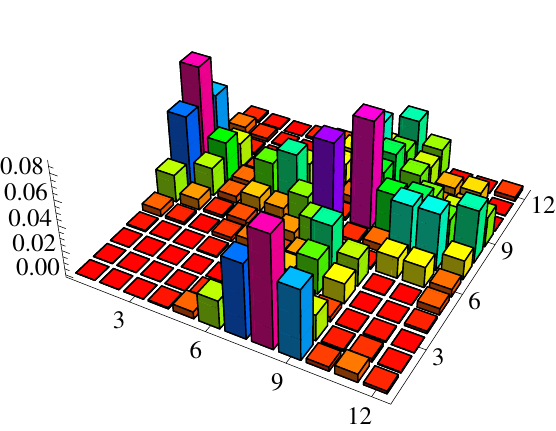} }
\label{Corr_Mob_t3}
}
\caption[]{ Two-photon correlation functions $\Gamma^{1,7}_{r,s}(n\tau)$ (Eq.~\eqref{Gammat}) for cylindrical array (top row) a) $n=0$, b) $n=1$, c) $n=2$, d) $n=3$ and for the M\"{o}bius strip array (bottom row) e) $n=0$, f) $n=1$, g) $n=2$, h) $n=3$. Here we set $\tau=1$, measure time in units of $g^{-1}$ and for clarity we re-scale the vertical axis by $\cos^{4(n-1)}(\theta)  \sin^4(\theta)$. Note the mirror flipping of the photon statistics at odd time steps in the  M\"{o}bius strip array on the bottom row.
}
\label{MobiusFig123}
\end{figure*}

In this section we discuss three particular waveguide implementations of our proposal: the cylindrical array, the M\"{o}bius strip array and the twisted circular array. For each implementation we diagonalize the local Hamiltonian, describe the geometry dependent permutation function and discuss the dynamics of a two-photon input state.

\subsection{Cylindrical Array}

The simplest implementation of our proposal is the cylindrical array presented in Fig.~\ref{LinearCircleArray}. Here, an equispaced one-dimensional (1D) waveguide array is curved around the surface of a cylinder with a large radius. We require the radius to be the largest length scale in the system, much larger than both the inter-waveguide spacing and the coherence length of the injected photons. Photons are injected into the input waveguides, which couple into the array via a series of directional couplers. The photons then evolve on the effective 1D waveguide array around the surface of the cylinder. After one traversal of the cylinder, the photons interact with the directional couplers where some photons may couple into the output modes to be detected. The remaining photons will then traverse the cylinder a second time and interact with the couplers. Depending on the coupling angle $\theta$, this process may occur for a large number of traversals of the cylinder. The relevant permutation is just the identity, 
$\mathbf{p} (j)=j$.

The equispaced 1D waveguide array has recently been the subject of intense theoretical~\cite{Agarwal} and experimental work~\cite{OBrienQW}. In particular, as the photon detectors were placed at the end of the 1D array, the experiment was only  able to measure the array at one particular interaction time ~\cite{OBrienQW}. In contrast the implementation presented here allows the 1D array to be observed at several discrete times with the same device. This ability allows the experiment to observe time-dependent properties of the CTQW.

We now consider the evolution of the photons in the cylindrical waveguide array. The effective local Hamiltonian of the array in Fig.~\ref{LinearCircleArray} is the 1D waveguide array Hamiltonian, 
\begin{eqnarray}
\hat{H}/\hbar &=& \omega \sum^{N}_{j=1} \hat{a}^{\dagger}_j  \hat{a}_j +g \sum^{N-1}_{j=1} \left( \hat{a}^{\dagger}_j  \hat{a}_{j+1}+ \hat{a}^{\dagger}_{j+1}  \hat{a}_j\right),
\label{LocalLinear}
\end{eqnarray}
where all waveguides have the same effective frequency $\omega$ and couple to their nearest neighbors at the rate $g$. This Hamiltonian is a special case of (\ref{GeneralHam}) with
\begin{eqnarray}
G_{j,k} = \omega  \delta_{j,k} + g(\delta_{j,k+1} + \delta_{j+1,k}).
\label{LinaerCouplingMat}
\end{eqnarray}
The evolution of the photons in the array can be determined from the Heisenberg picture mode operators (\ref{HeisenbergA}), which depend on the eigenvalues and eigenvectors of the coupling matrix (\ref{LinaerCouplingMat}). The eigenvalues and eigenvectors of this matrix are well known~\cite{Agarwal},
\begin{subequations}
\begin{eqnarray}
\lambda_j &=& \omega + 2g \cos \left(\frac{j \pi}{N+1} \right),\\
v_{j,k} &=& \sqrt{ \frac{2}{N+1}} \sin  \left(\frac{jk \pi}{N+1} \right).
\end{eqnarray}
\label{EigsLinear}
\end{subequations}
By substituting these relations into (\ref{psint}), we can find the evolution of the wavefunction of an arbitrary initial state in the device. Furthermore, using (\ref{EigsLinear}) it is straightforward to derive the two-photon correlation function $\Gamma^{j,k}_{r,s}  (n \tau)$ (\ref{Gammat}). The experimentally observable two-photon correlation function is shown in Fig.~\ref{Corr_Linear_t0}--\ref{Corr_Linear_t3} for several time steps. By injecting pairs of photons and detecting coincidences over many experimental runs, it is possible to build up several time steps of the two-photon correlation function. This is in contrast to current experiments, which can only measure the two-photon correlation function at a single time step~\cite{OBrienQW}.

\begin{figure*}[t]
\centering
\subfigure[]{
{\includegraphics[scale=0.7]{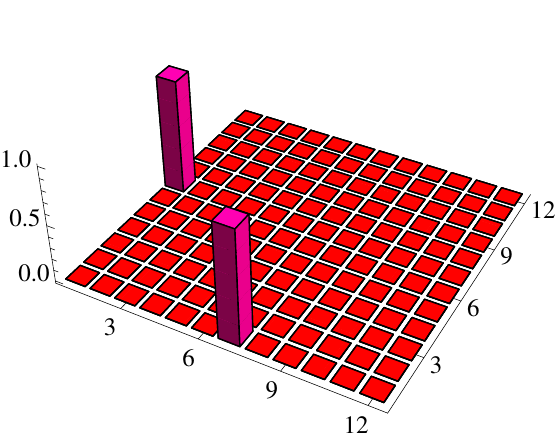}
 }
}
\subfigure[]{
{\includegraphics[scale=0.7]{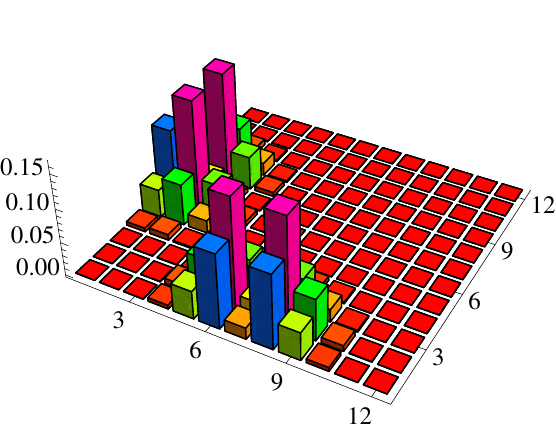}
 }
}
\subfigure[]{
{\includegraphics[scale=0.7]{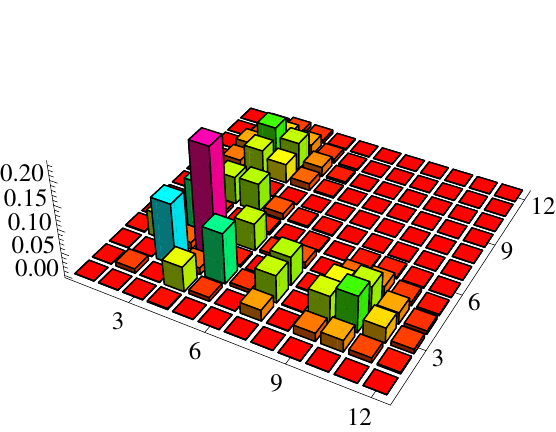}
 }
}
\subfigure[]{
{\includegraphics[scale=0.7]{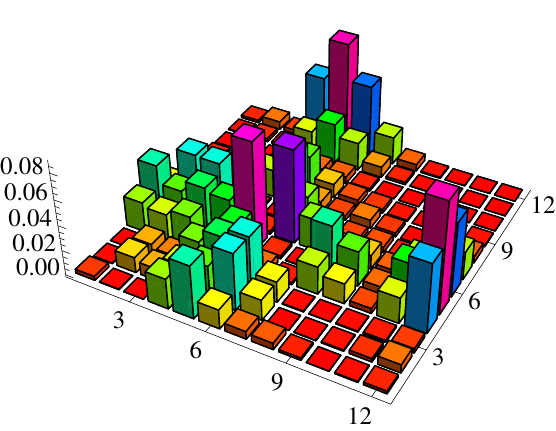}
 }
}
\subfigure[]{
{\includegraphics[scale=0.7]{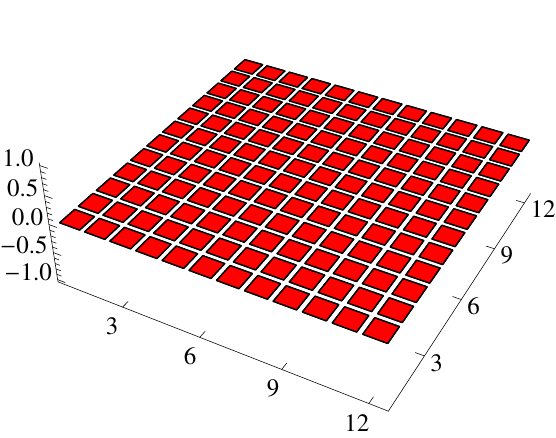}
 }
}
\subfigure[]{
{\includegraphics[scale=0.7]{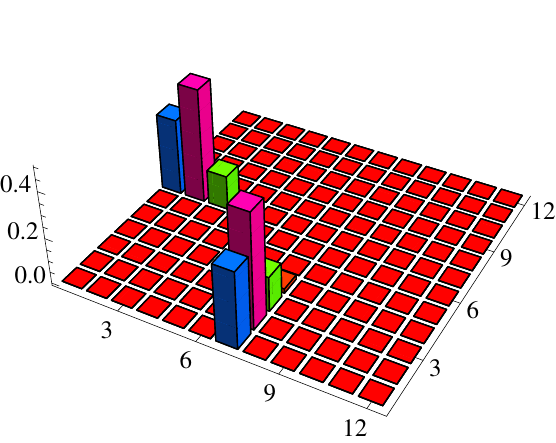}
 }
}
\subfigure[]{
{\includegraphics[scale=0.7]{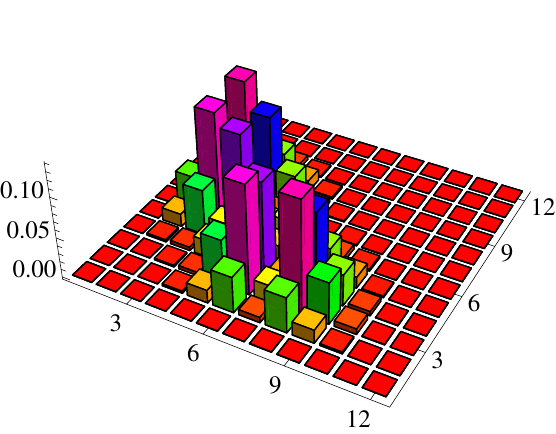}
 }
}
\subfigure[]{
{\includegraphics[scale=0.7]{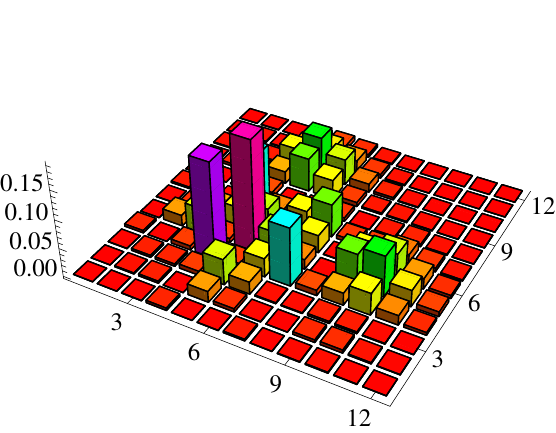}
 }
}
\subfigure[]{
{\includegraphics[scale=0.7]{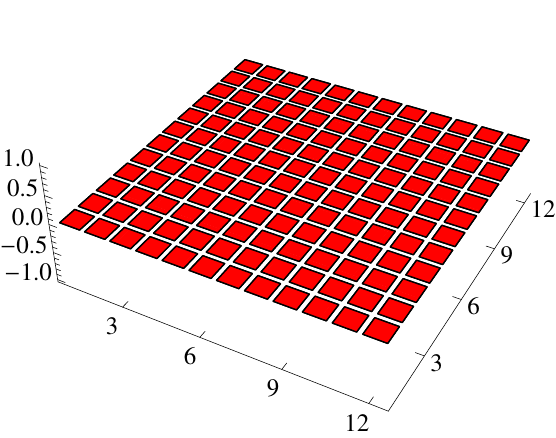}
 }
}
\subfigure[]{
{\includegraphics[scale=0.7]{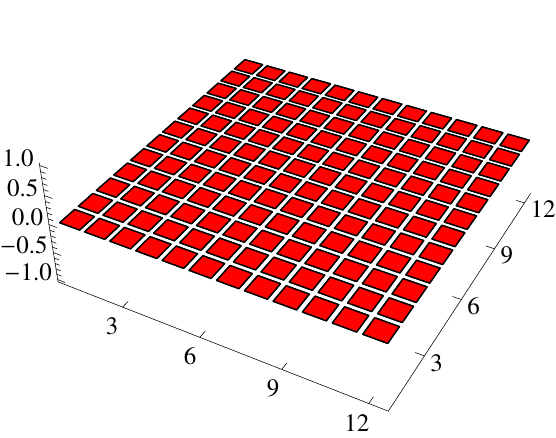}
 }
}
\subfigure[]{
{\includegraphics[scale=0.7]{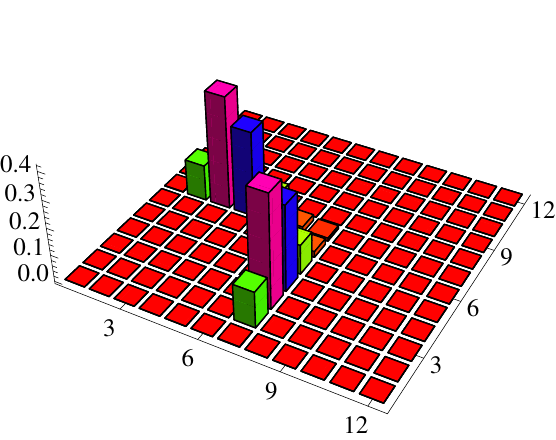}
 }
}
\subfigure[]{
{\includegraphics[scale=0.7]{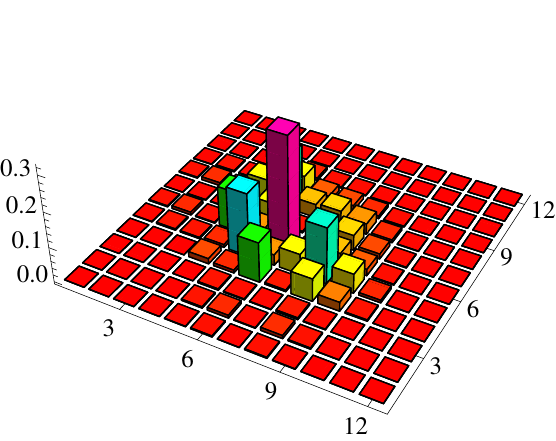}
 }
}
\caption[]{ 
 Two-photon correlation functions $\Gamma^{1,7}_{r,s}  (n \tau +n_d \tau)$ (Eq.~(\ref{Gammatdelayed})) for cylindrical array with delayed two-photon input. Top row: $n_d=0$ (simultaneous two-photon input) a) $n=0$ b) $n=1$ c) $n=2$ d) $n=3$. Middle row: $n_d=1$ (input the second photon after one traversal)
  e) $n=0$ f)  $n=1$  g) $n=2$ h) $n=3$.
 Bottom row: $n_d=2$ (input the second photon after two traversals)
  i) $n=0$ j) $n=1$  k) $n=2$ l) $n=3$
Here we set $\tau=1$, measure time in units of $g^{-1}$ and for clarity we re-scale the vertical axis by $\cos^{4(n-1)}(\theta)  \sin^4(\theta)$.
}
\label{CylindricalDelayed}
\end{figure*}

We also consider the case of two-photon input with delay in the cylindrical waveguide array. The time evolution of the two-photon correlation functions for three different delays are shown in Fig.~\ref{CylindricalDelayed}. The top row in Fig.~\ref{CylindricalDelayed} has simultaneous two-photon input, the second row has the second photon delayed by $\tau$, and the third row has the second photon delayed by $2\tau$. We see that states with different input delays have quite different statistics. Therefore by varying the delay, a variety of multimode entangled states can be prepared.

\subsection{M\"{o}bius Strip Array \label{MobiusSection}}

We now turn to the  M\"{o}bius strip array depicted in Fig.~\ref{Mobius_View1}. This device is intrinsically three dimensional, and outside the domain of conventional lithography. Similar to the cylindrical waveguide array, photons are injected in the input modes, which couple into the array via a series of directional couplers. The waveguide array then traces out a large loop, with a single twist, forming a M\"{o}bius strip. The twist in the loop implies that after a $2 \pi$ rotation, the first waveguide becomes the $N$-th waveguide and vice-versa. This geometrically induced permutation is described by the permutation function $\mathbf{p} (j)=N+1-j$. This permutation has the property $\mathbf{p}_2 (j)=j$, which implies that after a $4 \pi$ rotation, the $j$-th waveguide is mapped onto itself.

Assuming the radius of the strip is large, we can approximate the local Hamiltonian of the array as the 1D waveguide array Hamiltonian (\ref{LocalLinear}). Therefore the eigenvalues and eigenvectors for the  M\"{o}bius strip array are the same as the previous section. This approximation is only valid where the path lengths of the waveguides are identical, which is only approximately true for a strip of large radius. Experimentally, the path lengths will need to be adjusted to make the path lengths identical, as will be discussed in section~\ref{SecExperimentalIssues}.

The  M\"{o}bius strip device has a novel topology for CTQWs, whereby the walkers repeatedly evolve for a period of time then undergo a permutation. This topology is readily observable in the photon statistics, through the dependence of the correlation function (\ref{Gammat}) on the permutation function. The two-photon correlation for the M\"{o}bius strip is shown in Fig.~\ref{Corr_Mob_t0}--\ref{Corr_Mob_t3}, where we see that the strip topology induces a mirror reflection of the statistics at odd  time steps. Therefore, this device allows discrete time observation of a CTQW with time-dependent symmetries of the photon statistics.

\subsection{Twisted Circular Array}

\begin{figure*}[t]
\centering
\subfigure[]{
{\includegraphics[scale=0.7]{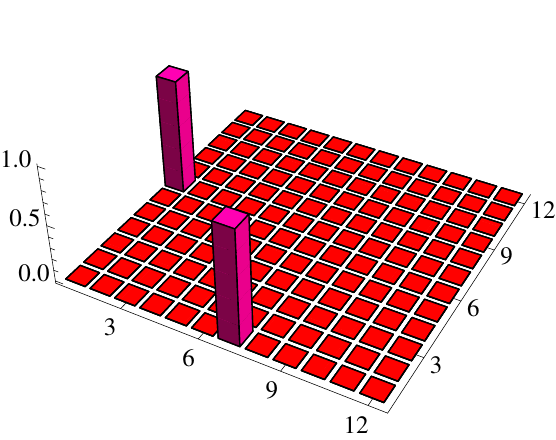}
 }
}
\subfigure[]{
{\includegraphics[scale=0.7]{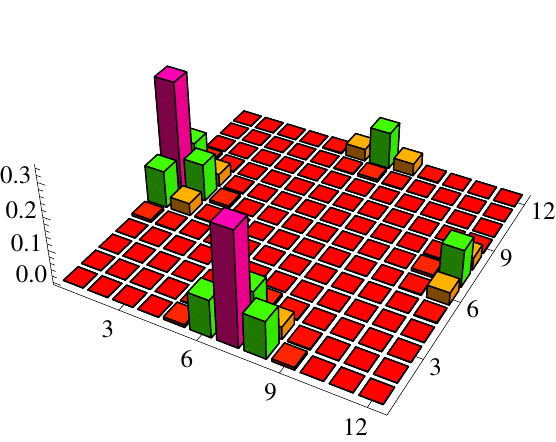} }
}
\subfigure[]{
{\includegraphics[scale=0.7]{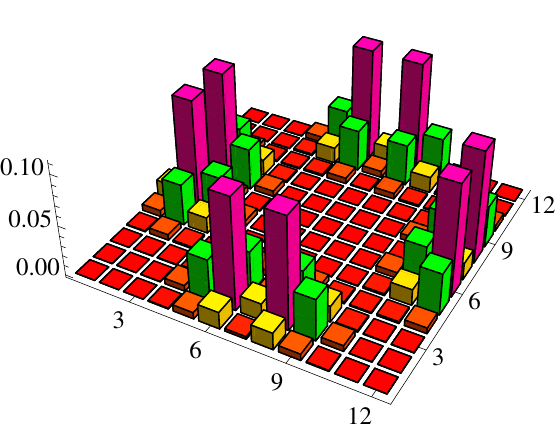} }
}
\subfigure[]{
{\includegraphics[scale=0.7]{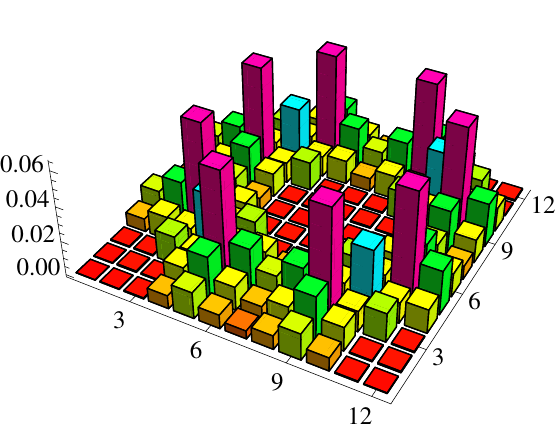} }
}
\subfigure[]{
{\includegraphics[scale=0.7]{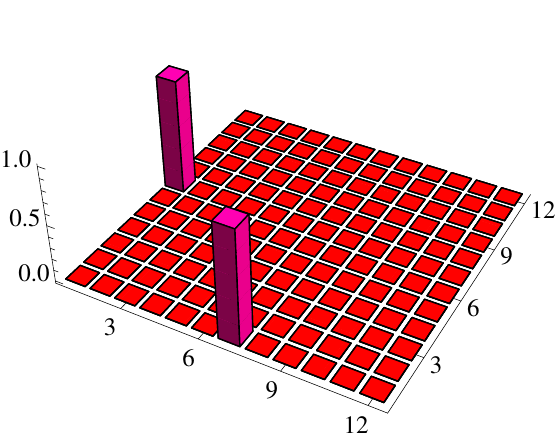}
 }
}
\subfigure[]{
{\includegraphics[scale=0.7]{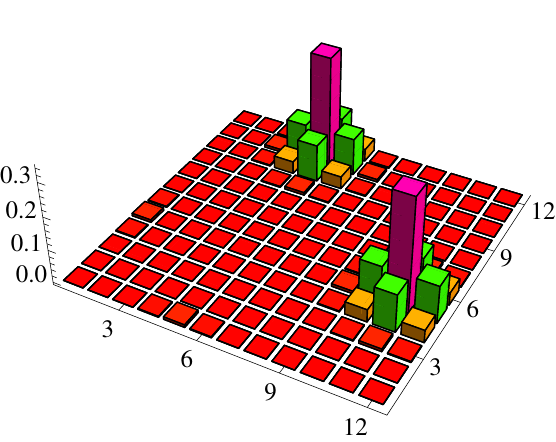} }
}
\subfigure[]{
{\includegraphics[scale=0.7]{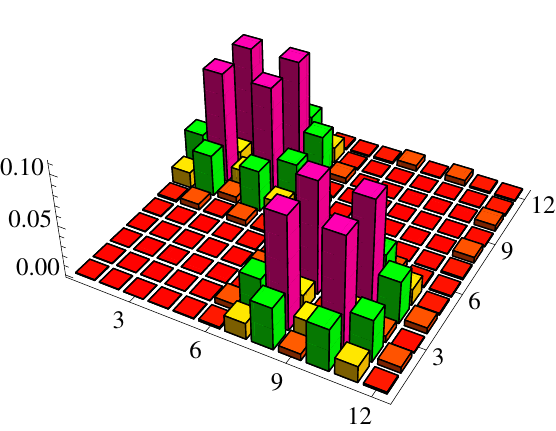} }
}
\subfigure[]{
{\includegraphics[scale=0.7]{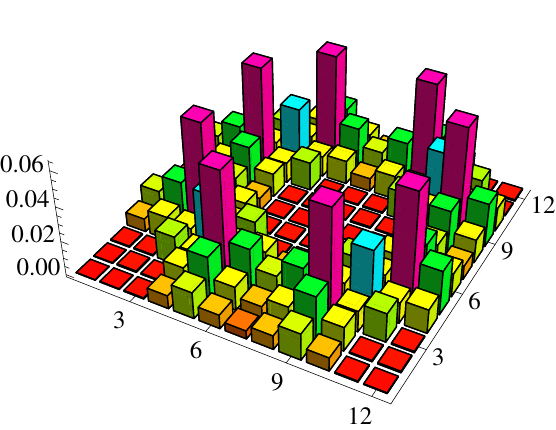} }
}
\caption[]{ Two-photon correlation functions $\Gamma^{1,7}_{r,s}(n\tau)$ (Eq.~\eqref{Gammat}) for circular array with no twists (top row) a) $n=0$, b) $n=1$, c) $n=2$, d) $n=3$ and for the twisted circular array (bottom row) 
e) $n=0$, f) $n=1$, g) $n=2$, h) $n=3$.
For clarity we re-scale the vertical axis by $\cos^{4(n-1)}(\theta)  \sin^4(\theta)$. On the twisted array (bottom row) there is a $4 \times 2\pi/12$ rotation each loop.
}
\label{TwistedCircleFig}
\end{figure*}

We now consider the twisted circular array depicted in Fig.~\ref{Lines_of_Revolution_Circle}. This array is a 3D generalization of the  2D  waveguide array recently experimentally studied~\cite{ThornPaper}. The array is formed by bending a 2D circular waveguide array into a large loop in the shape of a torus. By rotating the waveguides in the 2D array during the loop, it is possible to introduce a permutation of the modes. For example for $N=3$ waveguides, a $2\pi/3$ rotation during the loop causes the permutation, $\hat{a}_1 \rightarrow \hat{a}_2, \hat{a}_2 \rightarrow \hat{a}_3, \hat{a}_3 \rightarrow \hat{a}_1$.

The local Hamiltonian of the array, including the coupling between all non-adjacent modes, is,
\begin{eqnarray}\label{Hamiltonian}
H/\hbar = \sum^N_{\alpha=1} \sum^{N}_{n=1} g_{\alpha} \hat{a}^{\dagger}_n \hat{a}_{n+\alpha-1},
\end{eqnarray}
where due to the circular geometry, the subscript, $n+\alpha-1$, is modulo $N$, and due to symmetry, $g_{j> \frac{N}{2}+1} \equiv g_{N-j+2}$. Here the coupling matrix, $G$, is a circulant matrix defined by the  vector $(g_1, g_2, \dots, g_N)$. For example for $N=5$,
 \begin{equation}
G =
\begin{pmatrix}
g_1 & g_2 & g_3 & g_4 & g_5 \\
g_5 & g_1 & g_2 & g_3 & g_4 \\
g_4 & g_5 & g_1 & g_2 & g_3 \\
g_3 & g_4 & g_5 & g_1 & g_2 \\
g_2 & g_3 & g_4 & g_5 & g_1 \\
  \end{pmatrix} 
 =
\begin{pmatrix}
g_1 & g_2 & g_3 & g_3 & g_2 \\
g_2 & g_1 & g_2 & g_3 & g_3 \\
g_3 & g_2 & g_1 & g_2 & g_3 \\
g_3 & g_3 & g_2 & g_1 & g_2 \\
g_2 & g_3 & g_3 & g_2 & g_1 \\
  \end{pmatrix}  .   
\end{equation}
As $G$ is a circulant matrix, it can be diagonalized by the discrete Fourier transform (DFT). Using the DFT we find the eigenvalues and eigenvectors of $G$ are,
\begin{subequations}
\begin{eqnarray}
\lambda_j &=& \sum^N_{k=1}  g_{k}  e^{\frac{2 \pi i}{N} (j-1)(k-1)},\\
v_{j,k} &=& \frac{1}{\sqrt{N}} e^{-\frac{2 \pi i}{N}  (j-1)(k-1)}.
\end{eqnarray}
\end{subequations}
To our knowledge this is the first exact diagonalization of the circular waveguide array with non-nearest neighbor couplings.

The twisting of the array as it completes a loop causes a permutation of the modes. The simplest permutation is a $2 \pi/N$ rotation of the 2D circular array as it completes as single loop,
\begin{eqnarray}
\mathbf{p} (j) =
\begin{cases} 
 j+1 &\mbox{if } j \neq N \\
 1 & \mbox{if } j=N. 
\end{cases} 
\end{eqnarray}
After $N$ traversals of the loop, the modes return to their original order, $\mathbf{p}_N (j)=\mathbf{p} (j)$. The permutations from twisting the array leads to non-trivial  photon statistics. For example for $N=3$, a  $2 \pi/3$ rotation causes the correlation matrix to permute as,
 \begin{eqnarray}
\begin{pmatrix}
\Gamma^{j,k}_{1,1}  & \Gamma^{j,k}_{1,2}  & \Gamma^{j,k}_{1,3}   \\
\Gamma^{j,k}_{2,1}  & \Gamma^{j,k}_{2,2}  & \Gamma^{j,k}_{2,3}   \\
\Gamma^{j,k}_{3,1}  & \Gamma^{j,k}_{3,2}  & \Gamma^{j,k}_{3,3}   \\
  \end{pmatrix} 
&\rightarrow &
\begin{pmatrix}
\Gamma^{j,k}_{3,3}  & \Gamma^{j,k}_{3,1}  & \Gamma^{j,k}_{3,2}   \\
\Gamma^{j,k}_{1,3}  & \Gamma^{j,k}_{1,1}  & \Gamma^{j,k}_{1,2}  \\
\Gamma^{j,k}_{2,3}  & \Gamma^{j,k}_{2,1}  & \Gamma^{j,k}_{2,2}   \\
  \end{pmatrix} \nonumber
  \\
\rightarrow 
\begin{pmatrix}
\Gamma^{j,k}_{2,2}  & \Gamma^{j,k}_{2,3}  & \Gamma^{j,k}_{2,1}   \\
\Gamma^{j,k}_{3,2}  & \Gamma^{j,k}_{3,3}  & \Gamma^{j,k}_{3,1}   \\
\Gamma^{j,k}_{1,2}  & \Gamma^{j,k}_{1,3}  & \Gamma^{j,k}_{1,1}   \\
  \end{pmatrix}
  &\rightarrow &
\begin{pmatrix}
\Gamma^{j,k}_{1,1}  & \Gamma^{j,k}_{1,2}  & \Gamma^{j,k}_{1,3}   \\
\Gamma^{j,k}_{2,1}  & \Gamma^{j,k}_{2,2}  & \Gamma^{j,k}_{2,3}   \\
\Gamma^{j,k}_{3,1}  & \Gamma^{j,k}_{3,2}  & \Gamma^{j,k}_{3,3}   \\
  \end{pmatrix} .
 \end{eqnarray}
We see that all elements of the correlation matrix shift across and down one element, except those in the last row and last column which move to the first row and column. More complex permutations can be generated by multiple applications of this  $2 \pi/N$ rotation of the 2D circular array each loop.

The time evolution of the two-photon correlation matrix of a twisted circular waveguide array with twelve waveguides is shown in Fig.~\ref{TwistedCircleFig}. The array has a $4 \times 2\pi/12$ rotation each loop, which causes the permutation $\hat{a}_j \rightarrow \hat{a}_{j+4}$. Hence after three traversals of the torus, the waveguides return to their original ordering. The effect of the permutation on the twisted circular array correlation function  in Fig.~\ref{TwistedCircleFig} is more complex than those of the M\"{o}bius strip array in Fig.~\ref{MobiusFig123}. Here, the elements of the correlation matrix shift across and down four elements each time step, except the last four rows and columns, which wrap around to to the first four rows and columns. By changing the number of $2 \pi/N$ rotations of the 2D circular array each loop, a variety of multimode entangled states can be prepared.

\section{Experimental Issues \label{SecExperimentalIssues}}
We now consider the conditions required for experimental demonstration of our ideas.  We consider two possible implementations:  waveguide structures written
in glass blocks (typically boro-silicates) by the femtosecond laser direct-write (FLDW) technique~\cite{davis96}, 
or a loop of a suitably designed multicore fiber. 
The FLDW method, which has already been applied to a number of 3D quantum photonic problems~\cite{DTQW_Integrated,ThornPaper}, can in principle produce any of the structures described above including both the main loops and the input/output couplers. There are some practical challenges in writing waveguides with the same
properties at different depths within the glass block, but these can be significantly ameliorated using wavefront correction techniques~\cite{jes10}. On the other hand existing multicore fibers~\cite{zhu10} could be tapered to induce weak coupling between the different cores to create the circular and twisted circular arrays.  In this approach the input/output couplers would be spliced onto the multicore fiber to close the loops. This fiber based approach has the additional benefit of large loop radii as fibers of lengths of hundreds of metres are readily achievable.

Since the basic waveguide parameters---index contrast waveguide width, mode size, group index, coupling strength, etc,---are similar for these two approaches we consider only the FLDW system in detail.
We consider operation at a wavelength $\lambda=800$~nm to exploit standard spontaneous down-conversion photon sources and efficient silicon avalanche photodiode single photon detectors, 
but a configuration for $\lambda=1550$~nm is also realistic.  
We next demonstrate that competing limits associated with constraints on structure sizes,  photon loss due to bend-induced radiation, dispersion-induced pulse spreading, and coupler bandwidths permit a feasible working parameter space.

Using waveguide parameters comparable to those in previous experiments in quantum integrated photonics~\cite{marshall09,ThornPaper}, we assume a Gaussian index profile with maximum index contrast $\delta n=1.5\times 10^{-3}$, a $1/\mathrm{e}^2$ width $w=4.5$~\micron \; and a background index of $n_{bg}=1.44$.  Typical writing setups restrict the total device size to approximately $40$~cm dimensions, so we assume a loop radius of $R_c= 20$~cm and length $L\approx 2 \pi R_c$.  Standard results in optical waveguide analysis for Gaussian profile waveguides~\cite{snyderandlove} then yield a normalized frequency $V=2.32$ which is in the
single mode regime when $V<2.405$. For this loop radius the attenuation coefficient due to bend loss is $\gamma =6.8\times10^{-7} \mbox{cm}^{-1}$ which
implies a loss fraction of only 0.9\% after 100 transits around the loop.

We suppose the input photons are transform-limited with a pulse width of $T=20$~ps and a bandwidth
$\Delta \omega/(2 \pi)=$50~GHz.  The spatial extent of the pulses on input is then $l=v_g T=0.41$~cm.  Assuming
typical silica material dispersion, the net group velocity dispersion is $D=-150$~ps/nm/km so that the temporal width increases by less than $2\%$ after 100 transits around the loop.  The waveguide dispersion is even weaker and may be neglected. Therefore as the photons  can propagate many times around the loop with minimal loss and with a spatial pulse length always satisfying $l_p \ll L$, it is appropriate to treat the photons in the discrete fashion we have adopted in the analysis above. Furthermore, the bandwidth $\Delta\omega$ of the pulses corresponds to a wavelength bandwidth of $\Delta \lambda\approx 17$~pm and a relative bandwidth of $\Delta \lambda/\lambda=2.1\times10^{-5}$.  Over this
bandwidth a typical input/output coupler with waveguide separation $d = 10$~\micron \; shows a coupling variation of the same scale $\sim 2\times10^{-5}$, so that it is appropriate to assume a single directional coupler reflectivity for the photon pulses.

Due to the shape of the focal spot of the writing laser beam, the waveguides are typically slightly elliptical in cross-section.  One would arrange to launch the photons into one particular polarization axis of the waveguide, e.g.~the vertical.  As the waveguides move around each other and the coupling changes from left-right to up-down, there will be a small change in the coupling strength with position around the loop.  However the net behavior is still described by a single unitary matrix $U$ so there is no significant change to our analysis. 

Finally, we have also assumed the optical path lengths in each loop to be the same by tailoring the fan-out section at the directional couplers. Due to fluctuations in writing conditions and in the local properties of the glass block, this is difficult to accomplish directly.  However it is possible to post-tune the structures by additional laser exposure after characterization and so equalize the path lengths.

Therefore we have shown that the waveguide array devices considered in this paper are experimentally feasible in the near future. In particular we found a parameter regime where the photon loss due to bend-induced radiation, dispersion-induced pulse spreading, and coupler bandwidths have only a minor effect on the systems' dynamics. This justifies the assumptions made in the theoretical model in section~\ref{SecDescription}.

\section{Conclusion \label{SecConclusion}}

In summary, we have studied the dynamical evolution of CTQWs on several exotic waveguide arrays in  3D integrated photonics. Two of the primary benefits of the design are the ability to discretely observe CTQWs and the ability to inject additional photons during the quantum walk. We found that properties of the waveguide array topology are readily observable in the time-dependent photon correlations. By targeting time steps with particular topology induced symmetries or injecting  photons with a time delay it is possible to engineer a variety of multimode entangled states.

\acknowledgments
We thank Dr Graham Marshall and Thomas Meany for helpful suggestions on how to implement 
some of these structures.  This research was supported by the Australian Research Council Centre of Excellence for Ultrahigh bandwidth Devices for Optical Systems (project number CE110001018).
M.D. thanks  Professor Jingbo Wang of the Quantum Dynamics and Computation Research Group
at the University of Western Australia for her hospitality during the final preparation
of this manuscript.

\end{document}